\documentclass[sigplan,screen]{acmart}
\usepackage{microtype} 
\usepackage{tcolorbox}
\usepackage{listings}
\usepackage{xcolor}
\usepackage{subcaption}
\usepackage{multicol}

\AtBeginDocument{%
  }

\setcopyright{cc}
\setcctype{by}
\acmDOI{10.1145/3759425.3763390}
\acmYear{2025}
\copyrightyear{2025}
\acmISBN{979-8-4007-2148-9/25/10}
\acmConference[LMPL '25]{Proceedings of the 1st ACM SIGPLAN International Workshop on Language Models and Programming Languages}{October 12--18, 2025}{Singapore, Singapore}
\acmBooktitle{Proceedings of the 1st ACM SIGPLAN International Workshop on Language Models and Programming Languages (LMPL '25), October 12--18, 2025, Singapore, Singapore}
\received{2025-07-06}
\received[accepted]{2025-08-08}

\lstdefinelanguage{TypeScript}{
  sensitive=true,
  alsoletter={\#},                       
  morekeywords=[1]{break,case,catch,class,const,continue,debugger,default,delete,do,
    else,export,extends,false,finally,for,function,if,import,in,instanceof,new,null,
    return,super,switch,this,throw,true,try,typeof,var,void,while,with,yield,async,await},
  morekeywords=[2]{type,interface,implements,public,private,protected,readonly,abstract,
    override,declare,namespace,module,enum,keyof,typeof,infer,as,asserts,satisfies},
  morekeywords=[3]{any,unknown,never,number,string,boolean,bigint,symbol,object,
    undefined,null,void},
  morekeywords=[4]{Array,ReadonlyArray,Record,Partial,Required,Pick,Omit,Exclude,Extract,
    NonNullable,Parameters,ConstructorParameters,ReturnType,InstanceType,ThisType,
    Promise,Map,Set,WeakMap,WeakSet},
  morecomment=[l]{//},
  morecomment=[s]{/*}{*/},
  morestring=[b]',                         
  morestring=[b]",                         
  morestring=[b]`                          
}
\lstdefinestyle{acmcode}{
  basicstyle=\footnotesize\ttfamily, 
  columns=fullflexible,              
  keepspaces=true,
  showstringspaces=false,
  breaklines=true, breakatwhitespace=true,
  frame=none,
  numbers=none, numberstyle=\scriptsize, numbersep=6pt,
  captionpos=b,
  aboveskip=0.6\baselineskip, belowskip=0.3\baselineskip,
  moredelim=[is][\color{red}\bfseries]{@R@}{@R@},
  moredelim=[is][\color{blue}\bfseries]{@B@}{@B@}
}

\pdfoutput=1
\begin{document}

\title{Position: Vibe Coding Needs Vibe Reasoning: Improving Vibe Coding with Formal Verification}


\author{Jacqueline Mitchell}
\authornote{Both authors contributed equally to this research.}
\authornote{Corresponding Author}
\affiliation{%
  \institution{University of Southern California}
  \city{Los Angeles}
  \country{USA}}
\email{jlm41510@usc.edu}

\author{Yasser Shaaban}
\authornotemark[1]
\affiliation{%
  \institution{Workato}
  \city{Palo Alto}
  \country{USA}}
\email{yasser.shaaban@msn.com}


\renewcommand{\shortauthors}{Mitchell and Shaaban}
\settopmatter{printacmref=true}

\begin{abstract}
``Vibe coding'' --- the practice of developing software through iteratively conversing with a large language model (LLM) --- has exploded in popularity within the last year.
However, developers report key limitations including the accumulation of technical debt, security issues, and code churn to achieve satisfactory results.
We argue that these pitfalls result from LLMs' inability to reconcile accumulating human-imposed constraints during vibe coding, with developers inadvertently failing to resolve contradictions because LLMs prioritize user commands over code consistency.
Given LLMs' receptiveness to verification-based feedback, we argue that formal methods can mitigate these pitfalls, making vibe coding more reliable.
However, we posit that integrating formal methods must transcend existing approaches that combine formal methods and LLMs.
We advocate for a side-car system throughout the vibe coding process which: (1) \emph{Autoformalizes} specifications (2) Validates against targets, (3) Delivers \emph{actionable} feedback to the LLM, and (4) Allows intuitive developer influence on specifications.
\end{abstract}



\maketitle

\section{Introduction}

Modern LLMs have made vibe coding --- writing software by conversing with an LLM --- an appealing new workflow,
granting developers the ability to rapidly prototype and refine code by prompting an LLM \cite{edwards2025vibe, sapkota2025vibe}.
In the context of software development (and developing a mature codebase), vibe coding is a \emph{long-range} iterative venture, guided by human feedback.
Over time, the user's requirements accumulate as natural language constraints on the system.
These constraints may be inconsistent (e.g., conflicting design goals) due to human error, evolving goals, or contradictory instructions from the user.
In fact, developers have reported dealing with accumulated technical debt (and security issues) that compound over time \cite{sewell2024ai, gleason2025chorus}.
In extreme cases, developers report falling into the so-called ``pit of despair,'' a state where starting over is easier than untangling inconsistencies in the code \cite{sewell2024ai}.
LLM-generated code may also contain security flaws:
a recent repository-level evaluation of 318 benchmark programs shows that the strongest model generates secure and correct code at a rate of only 28\% \cite{DBLP:journals/corr/abs-2504-21205}.

Formal verification is increasingly used to check security properties in LLM-generated code \cite{DBLP:journals/corr/abs-2412-05098, yao2025real, DBLP:journals/corr/abs-2412-14841, DBLP:conf/ease/KavianKKFG24}.
It can automatically assess safety and intended behavior, making it attractive for vibe coding.
However, the ability to verify specifications against LLM-generated code is only one part of the battle.
The burden of writing and maintaining application-specific specifications (i.e., beyond simple checks, such as making sure the code type-checks) falls upon developers and grows more burdensome with rapid iteration on the code.
Furthermore, even prior to vibe coding, industry reports show that it is difficult to get developers to adopt static analysis in their workflows \cite{DBLP:journals/cacm/SadowskiAEMJ18, DBLP:journals/cacm/DistefanoFLO19, DBLP:journals/corr/abs-2310-00205, DBLP:conf/icse/JohnsonSMB13}.

This paper argues that formal methods can improve vibe-coded software, provided the integration is developer-first. 
We propose \textbf{Vibe Reasoning}: a system that autoformalizes application-specific verification targets, verifies them with the lightest effective techniques, provides feedback that leads to actionable fixes, and keeps developers in control in a way where human involvement is collaborative, instead of an overhead. 
Unlike typical autoformalization or full-program verification, it iteratively and automatically checks only critical invariants.
\section{Challenges in Vibe Coding}

This section outlines the vibe coding process in more detail, emphasizing the key challenges.
\setlength{\textfloatsep}{10pt}
\begin{figure}
    \centering
    \includegraphics[width=\linewidth]{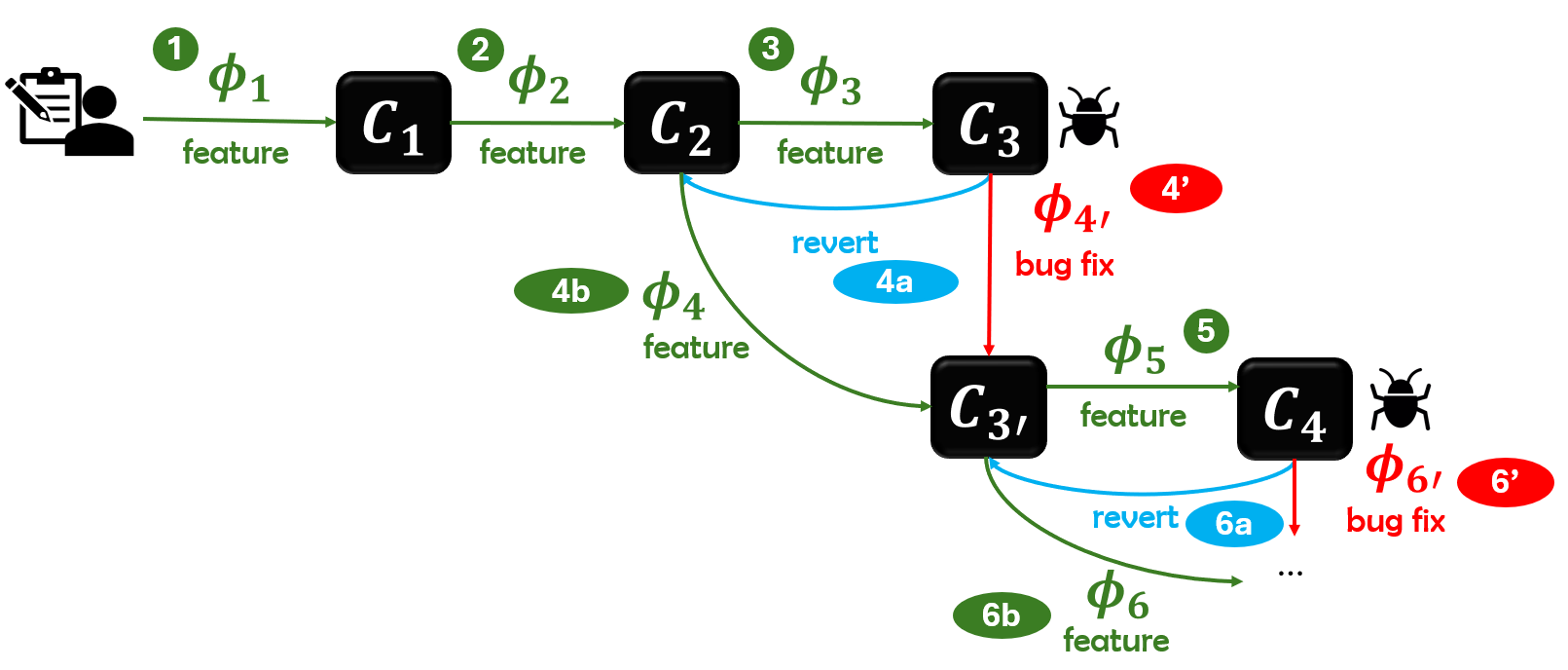}
    \caption{Example of a Vibe Coding Workflow}
    \label{fig:vibe-coding}
\end{figure}
Figure~\ref{fig:vibe-coding} depicts a possible vibe coding workflow.
While vibe coding, developers typically: (1) request new features or revisions (e.g., refactoring) or (2) when they find bugs or misalignment with their intent, either ask the LLM to fix them or revert to a prior version and continue.
The user's demands accumulate in the form of natural language constraints $(\phi_0, \phi_1, \ldots)$.
Over time, these constraints may drift or conflict, leaving the code in an inconsistent, buggy, and complex state, as reported by developers \cite{sewell2024ai, gleason2025chorus}.  
Common complications include: \textbf{(1) Constraint Inconsistency}, where a new feature silently contradicts existing behavior;
\textbf{(2) Partial-propagation Bugs}, where a change (e.g., in a schema or input-sanitization) is applied in one module, but not others, leading to errors (e.g., crashes or data-corruption);
\textbf{(3) State-machine Divergence}. For example, when a UI or protocol logic acquires new states or transitions in one part of the codebase, but the transitions are not globally updated, leading to crashes or illegal states; 
\textbf{(4) Duplication Debt}, where multiple LLM-generated variants of a routine evolve separately, ballooning technical debt.

These patterns mirror technical debt patterns in machine learning-based coding systems, including entanglement \\ (changing any input affects all outputs), hidden feedback loops, undeclared consumers (components that silently depend on changed behavior), and correction cascades (multiple versions of similar code evolve independently), described by Sculley et al. \cite{sculley2015hidden}.
In vibe coding, these patterns are amplified by the rapid development style and the developer’s limited visibility of the generated code’s interdependencies.
Vibe coding also compresses the development cycle:
projects that once required teams can now be done solo, removing review-driven guardrails and other implicit constraints. The solo developer must choose when and how to refactor (e.g., apply Single Responsibility Principle) while managing growing branches that hide latent conflicts and hard-to-fix bugs.

Balancing developer intent, evolving constraints, and codebase consistency demands systematic planning and reconciliation, capabilities which LLMs currently lack.
Kambhampati et al. \cite{DBLP:conf/icml/KambhampatiVGVS24} argue that LLMs optimize for local plausibility (pattern completion) rather than combinatorial planning needed for global consistency, so they struggle to reconcile evolving constraints over long horizons.
This results in \textbf{constraint-reconciliation decay}: as sessions lengthen and dependencies accumulate, success rates fall.
This is supported by several studies:
Multi-turn coding benchmarks show that performance steeply declines compared to single-turn tasks, dominated by failures to globalize changes across helpers and state \cite{DBLP:journals/corr/abs-2504-21751}.
Repository-level evaluations reveal similar cross-file integration and dependency breaks \cite{DBLP:journals/corr/abs-2412-00535,DBLP:journals/corr/abs-2406-11927}.
Success of agentic systems on static benchmarks (SWE-bench Verified) do not carry over to dynamic, evolving tasks (SWE-bench-Live) \cite{DBLP:journals/corr/abs-2506-17208,DBLP:journals/corr/abs-2505-23419}.
Further, relying on expanded ``memory'' is also insufficient; 
benchmarks show that long-term recall is unreliable and degrades under complexity \cite{DBLP:journals/corr/abs-2410-10813, DBLP:journals/corr/abs-2506-21605}, accumulating contradictions rather than ensuring consistency.

\section{Why Formal Methods in Vibe Coding?}
Despite the difficulties LLMs face with planning, they readily incorporate explicit, well-structured human feedback \cite{DBLP:conf/icml/KambhampatiVGVS24}.
In Figure~\ref{fig:vibe-coding}, at each iteration, the user provides feedback in the form of natural language.
Generally, the LLM adjusts accordingly (possibly in an ad-hoc manner) to immediate feedback.  
The task of generating this feedback falls upon the user, who must manually test or inspect the code, and triage what to test and when, which becomes cumbersome.
Offloading the burden to LLM-written tests is not ideal either.
Test suites must evolve with the code, and the developer must ensure that the tests reflect their intent and are consistent with the codebase.
Achieving high coverage often requires many tests whose maintenance can be more complex than that of the codebase itself. 
Running the tests can take a long time, and even a high pass rate cannot guarantee the absence of important classes of bugs.

Formal methods can guarantee the absence of bugs, formalize design constraints, and model and verify system-wide invariants and dependencies.
This makes their integration into vibe coding compelling, especially given the systemic nature of ML technical debt \cite{sculley2015hidden}.
Many existing works combine LLMs and formal methods, highlighting the ability of LLMs to effectively incorporate feedback from a verification engine.
The next subsection classifies these approaches and argues for the need for a developer-first integration that transcends existing systems.

\subsection{Type I and Type II Systems}
Work at the intersection of formal methods and LLMs for synthesis can largely be classified into two types.
\begin{figure}
    \centering
    \includegraphics[width=\linewidth]{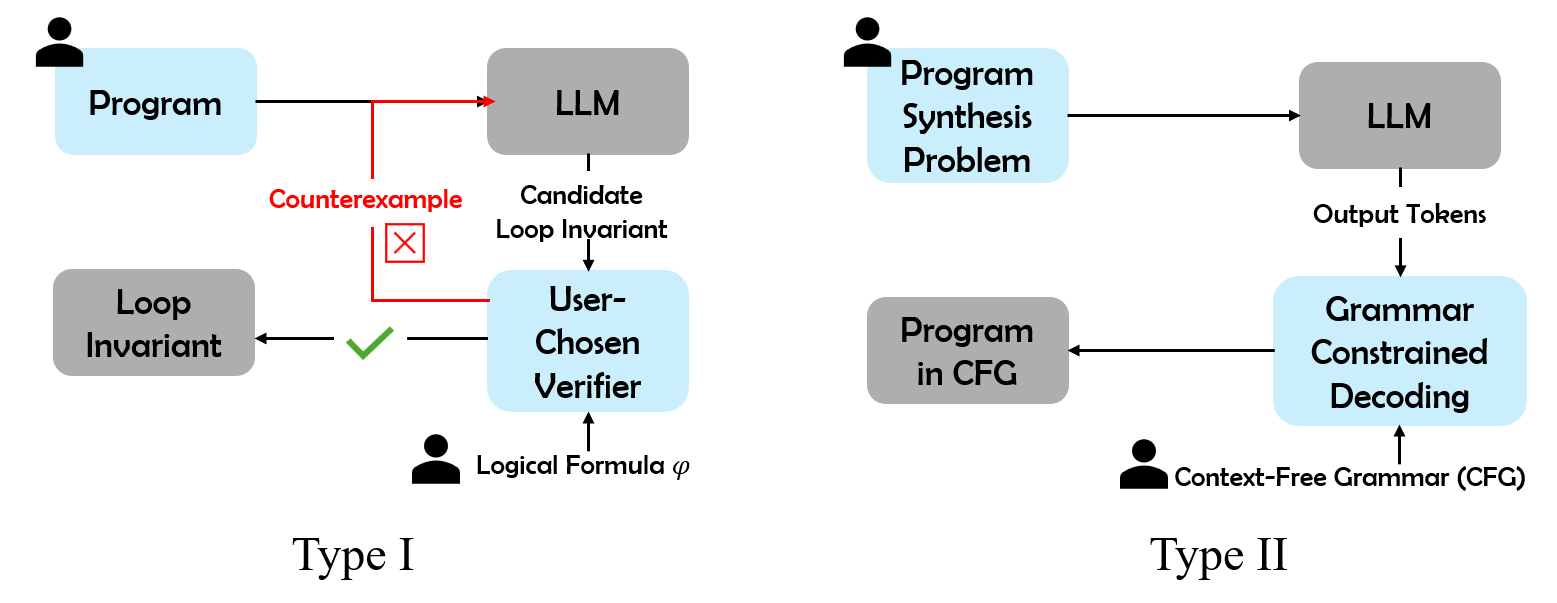}
    \caption{Examples of Type I and Type II Systems}
    \label{fig:t1vst2}
\end{figure}
\textbf{Type I} systems (Figure~\ref{fig:t1vst2}, left) use formal methods to filter out solutions that do not satisfy a certain specification, and provide feedback (e.g., a counterexample) to the LLM, iterating until a valid solution is reached
(e.g., loop-invariant synthesis with formal tools in the loop \cite{DBLP:conf/emnlp/ChakrabortyLFLM23, DBLP:conf/kbse/WuC0W0M24, DBLP:journals/corr/abs-2412-10483, DBLP:journals/corr/abs-2311-10483, DBLP:journals/corr/abs-2311-07948}, static analysis-guided repair \cite{DBLP:conf/aaai/OrvalhoJM25}).
\textbf{Type II} systems (Figure~\ref{fig:t1vst2}, right) use formal methods to post-process LLM outputs so that it satisfies specifications by construction
(e.g., constrained decoding \cite{DBLP:conf/nips/ParkWBPD24, DBLP:journals/corr/abs-2502-05111, DBLP:journals/corr/abs-2504-09246}, program completion \cite{DBLP:conf/sigsoft/0003X023}).

Both systems rely on humans to supply formal methods components (e.g., specifications). 
In a vibe coding workflow, verification targets may need to evolve over time; developers must decide how to convey the feedback to the LLM (Type I) or integrate post-processed results into the codebase (Type II).
Then, the user must ensure that the code remains in a consistent state, which may be very burdensome.
Thus, we advocate for a system (\textbf{Type III}), which incorporates formal methods in a way such that \emph{human involvement becomes collaborative, rather than an overhead cost}.

\section{The Type III Vibe Reasoning Trifecta}
We specify an autonomous (agentic \cite{sapkota2025vibe}) system centered on developer ease and collaboration.
\begin{figure*}
    \centering
    \begin{subfigure}{0.4\textwidth}
        \centering
        \includegraphics[width=\linewidth]{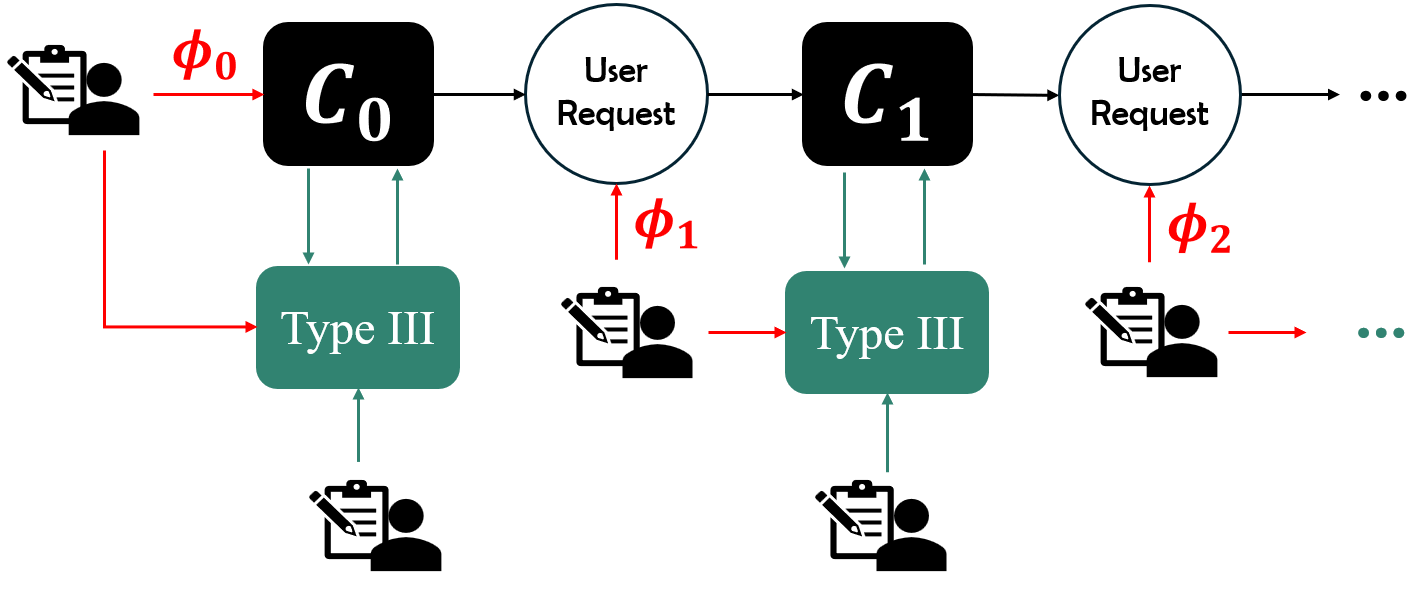}
        \caption{The Vibe Reasoning Side-Car Architecture}
        \label{fig:sidecar}
    \end{subfigure}\hfill      
    \begin{subfigure}{0.55\textwidth}
        \centering
        \includegraphics[width=\linewidth]{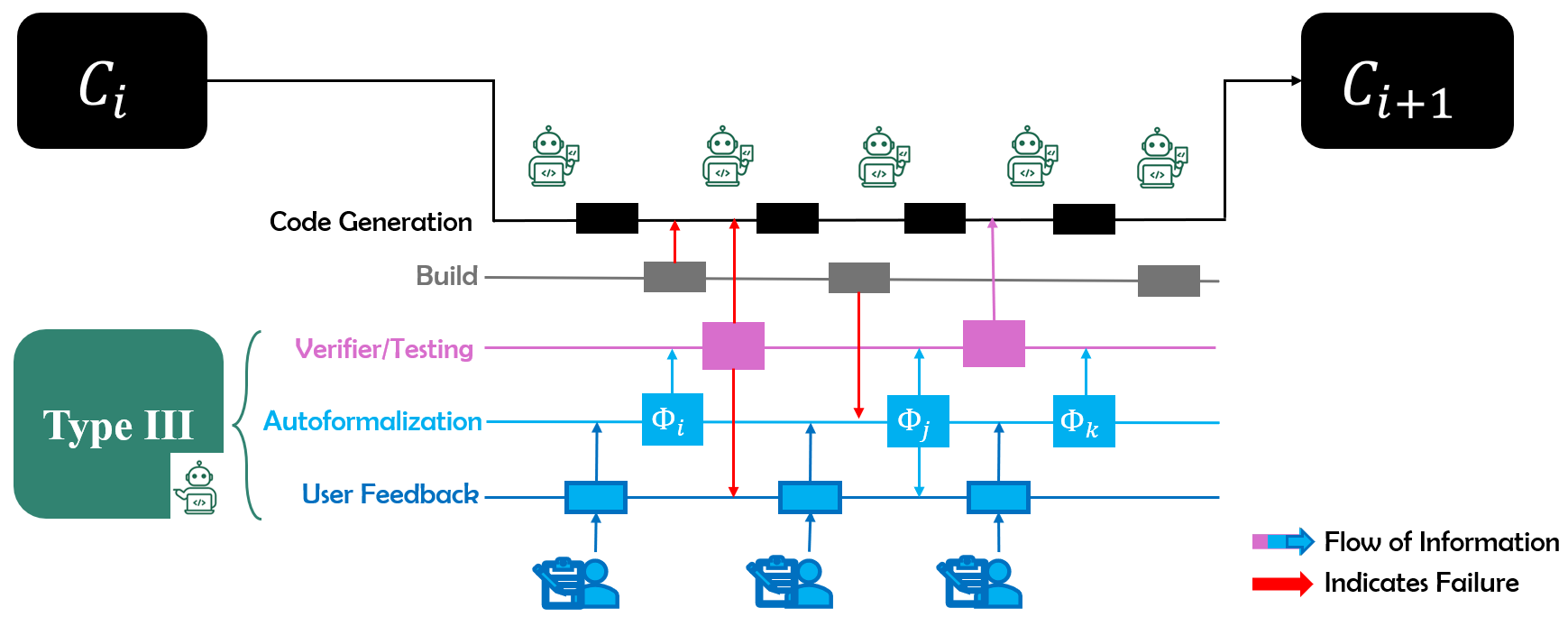}
        \caption{Type III Component}
        \label{fig:typethree}
    \end{subfigure}
    \caption{Type III: Vibe Reasoning}
    \label{fig:vibereasoning}
\end{figure*}
Figure~\ref{fig:sidecar} shows the high-level Vibe Reasoning loop, where the system runs alongside vibe coding (abstracted from Figure~\ref{fig:vibe-coding}).
Figure~\ref{fig:typethree} describes the Type III side-car,
which manages verification, autoformalization, and user-feedback integration.
Each block in Figure~\ref{fig:typethree} denotes an event (e.g., requesting user feedback on a verification target or autoformalizing a specification).
Red arrows represent failures (e.g., failure to verify), and communicate that information to various parts of the system.
The structure of this section follows the core components of the side-car: \textbf{(1) Autoformalizing Specifications}, \textbf{(2) Continuous Verification}, and \textbf{(3) Integration} (of human and verifier feedback).

\subsection{Autoformalizing Specifications}
The goal is to convert developer intent and good coding practices (e.g., maintainability and security) into formally verifiable specifications.
Some specifications are universal and tool-discoverable, including exhaustive variant handling (e.g., TypeScript's \texttt{assertNever}, Kotlin sealed classes), Effect-\\Dependency soundness (e.g., React's \texttt{exhaustive-deps} linter), and Parameterized SQL Enforcement (e.g., CodeQL, which detects unescaped concatenation).
The system should be able to apply these checks as needed.

Two key challenges in autoformalization are \textbf{(1)} \textbf{Creating high-quality \emph{application-specific} verification targets} aligned with explicit (e.g., a specification that guarantees the absence of a  discovered bug) and implicit (e.g., inventory in an e-commerce system is never negative) developer intent to limit technical debt and security risks.
\textbf{(2)} \textbf{Managing the constraints accumulated over time}:
new specifications may contradict earlier ones, or the number of specifications may be unwieldy.

\textbf{\emph{Promising Directions:}}
LLM-based autoformalization for verification shows promise \cite{DBLP:journals/corr/abs-2410-10135, DBLP:journals/corr/abs-2301-02195, DBLP:conf/nips/LiWLWZYM24, DBLP:conf/emnlp/ZhangQF24}, across a variety of application domains.
In the context of vibe coding, the goal is to map explicit and implicit intent (in the form of \emph{natural language}) into formally verifiable or checkable specifications.
Recent works suggest possible directions for achieving high-quality autoformalizations with LLMs.
Internal Coherence Maximization \cite{wen2025unsupervised} is an unsupervised search technique that selects mutually predictable and logically self-consistent LLM responses.
Another work uses probabilistic consensus for response selection and argues that unanimous consensus amongst several independent LLMs can serve as proxies for ground truth \cite{DBLP:journals/corr/abs-2411-06535}.
Zhou et al. demonstrate that an LLM can generate high-quality tasks via Code-as-Task bundles which pair natural language goals with an executable verifier function that can automatically validate task completion and provide feedback to improve model performance \cite{DBLP:journals/corr/abs-2506-01716}. 
For \textbf{Type III}, future work should explore if ensembles of LLMs can propose and select formal verification targets beyond test suites and if they can maintain the verification environment over time (e.g., deciding to keep or retire specifications) and consider feedback from failed builds (Figure~\ref{fig:sidecar}).

\textbf{\emph{Human Collaboration:}} Of course, LLM-generated specifications may not fully align with developer intent.
Thus, we advocate for a human-in-the-loop component.
Prior work shows improved trust in users when they participate in verification.
VeriPlan \cite{DBLP:conf/chi/LeePWZM25} lets end-users approve logical rules (in natural language) and choose which to enforce as hard or soft constraints.
For vibe coding, the system could present specifications in natural language with an explanation of their impact on the codebase and trade-offs, then let users approve, reject, or request changes.
This could address the second challenge: users can resolve conflicting design requirements by selecting specifications that meet their needs.

\subsection{Continuous Verification}
Once selected, specifications are used as verification targets throughout the vibe coding process, to be validated by appropriate tools (chosen by the system).
The system must decide which tools to use to verify the specifications.

\textbf{\emph{Challenge:}} A key challenge in continuous verification is that the scale at which verification is used must match development pace.  
Verification techniques that are too slow may be undesirable.
The system must make trade-offs by focusing on core components, preferring lightweight analyses when possible, and prioritizing the most important targets.

\textbf{\emph{Promising Directions:}} Agentic techniques have been shown to be effective at external tool selection \cite{DBLP:journals/corr/abs-2410-22457, DBLP:conf/emnlp/KongRC0BSQHM0ZZ24, DBLP:journals/corr/abs-2502-11705, DBLP:journals/corr/abs-2502-04644}.
The system could leverage these techniques, use knowledge about the trade-offs of different verification techniques, or rewrite components of code to be more verification-friendly.

\textbf{\emph{Human Collaboration:}} Users may wish to weigh in on the cost-benefit of verifying specific targets.
The system could explain benefits in natural language and allow users to continue or stop verification (e.g., when the runtime is too high) and to request alternative specifications when the verification conflicts with their pace.

\subsection{Integration}
The last tenet is effectively integrating feedback from a verifier to the LLM.
Typically, this is done by providing the feedback to the LLM using prompts, which has been shown to be generally effective in Type I systems.
If the LLM is unable to fix the code to satisfy a certain specification, Type II techniques may be invoked; when applicable, 
the system could select and apply a Type II technique to ensure that the generated component satisfies the specification.

\textbf{\emph{Challenge:}} Incremental modifications to the codebase are not guaranteed to be fool-proof.  
Edits made to satisfy one specification may regress another.

\textbf{\emph{Promising Directions:}} The side-car could incrementally incorporate verifier feedback as part of the compile/build process (Figure~\ref{fig:typethree}), ensuring accumulated constraints (the specifications) are enforced. 

\textbf{\emph{Human Collaboration:}} To involve the developer, the system could present proposed changes, with natural-language explanations, for approval or rejection. 
If integration to satisfy a specification takes too long, the user may relax it to a soft constraint, to steer development without enforcing a strict requirement.
\section{A Proof-of-Concept Type III}

We describe a simple Type III side-car that mitigates common TypeScript vibe coding errors that can cause state-machine divergence and error propagation as code evolves over time. 
The side-car integrates with a coding agent, triggering incremental analysis that proposes the introduction of new formalizations, currently focused on syntactic checks for a few patterns which can flag future bugs, that get checked via compiler.
The goal is to incrementally steer development such that as many bugs are detected at compile time as possible.

To illustrate the bugs we aim to prevent, consider the toy example in Listing~\ref{lst:lstex}, ignoring the \textcolor{red}{\textbf{red}} code for now.
Suppose that \textcolor{blue}{\texttt{`shipped'}} and \textcolor{blue}{\texttt{`cancelled'}} (and the associated code in \textcolor{blue}{\textbf{blue}}) were added via LLM edits.
Assuming that \texttt{updateOrderUI} calls \texttt{OrderBadge}, the LLM edits silently lead to two bugs:
(1) \texttt{OrderBadge} lacks a case for \texttt{`shipped'}, so shipped orders are not reflected in the UI.
(2) \texttt{processOrder} lacks a \texttt{`cancelled'} action, so the UI could indicate that an order has been cancelled, while no cancellation occurs.

{\scriptsize
\begin{lstlisting}[language=TypeScript,style=acmcode,
  caption={Toy function with exhaustive handling},label={lst:toy}, escapeinside={(*@}{@*)}, label={lst:lstex}]
export type OrderStatus = 'pending' | 'paid' | @B@'shipped'@B@ | @B@'cancelled'@B@

// orderProcessor.ts
function processOrder(order: Order){
  updateOrderUI(order.status);
  switch(order.status) {
    case 'pending': return sendPaymentReminder(order);
    case 'paid': return scheduleShipping(order);
    @B@case 'shipped': return sendNotification(order);@B@
    @R@// side-car: exhaustive guard@R@
    @R@default: return assertNever(order.status);@R@
  }
}

// orderUI.tsx
function OrderBadge({ status }: { status: OrderStatus }) {
  switch (status) {
    case 'pending': return <Badge color="Y">Pending</Badge>;
    case 'paid': return <Badge color="G">Paid</Badge>;
    @B@case 'cancelled': return <Badge color="Gr">Cancelled</Badge>@B@;
    @R@case 'shipped': return <Badge color="B">Shipped</Badge>@R@
  }
}
\end{lstlisting}
}
We now outline the component for Autoformalization (Figure~\ref{fig:typethree}), specialized for these kinds of TypeScript bugs.
Autoformalization uses LLM-generated specifications following predefined templates.
Verification is lightweight, relying on syntactic checks and successful compilation.
The specifications can also act as ``soft'' constraints, used as suggestions provided to the LLM to steer engineering practices, in hopes of revealing future bugs at compile time.

We focus on four autoformalization targets, which take the form of templates that are generated by an LLM.  
The templates are composed of two pieces: the \textbf{Scope} and the \textbf{Spec}. The Spec describes the desired properties, while the Scope describes which parts of the code are being considered as part of the Spec. 
The four targets are as follows:

\textbf{(1) Exhaustive Switch:} 
The Scope is a switch statement \texttt{S}, described by its location in the code, union type \texttt{UnionName} being switched over, and which cases are covered, \texttt{Cases}.
The Spec states that either $\mathtt{Cases} = \mathtt{Members(UnionName)}$ or \texttt{S} has a \textbf{\texttt{default}} containing \textbf{\texttt{assertNever}} and compilation succeeds.
\textbf{\texttt{assertNever}} (shown below) turns ``exhaustive handling'' into a compile-time check. If control reaches it with anything other than \texttt{never}, the TypeScript compiler errors immediately.

{\scriptsize
\begin{lstlisting}[language=TypeScript,style=acmcode,]
export function assertNever(x: never): never {
  throw new Error(`Unexpected case: ${JSON.stringify(x)}`);
}
\end{lstlisting}
}

The goal of Exhaustive Switch is to prevent silent failures when union types are extended, to enforce exhaustive handling of union types across the codebase, and ensure that future buggy variants fail quickly.
This spec is capable of detecting the bugs in the code of Listing~\ref{lst:lstex}. 

\textbf{(2) Discriminated Union:} 
The Scope is an area of control flow characterized by comparisons on a string-valued variable $t$ belonging to an object type and the observed values of $t$.
The Spec states that a union type \texttt{UT} should be created with the observed values of $t$ and that the string comparisons should be replaced with a \textbf{\texttt{switch}} statement over \texttt{UT}.
For example, in Listing~\ref{lst:lstex2}, $t$ is \texttt{action.type} and the observed values are \texttt{`ADD\_TODO'}, \texttt{`REMOVE\_TODO'}, and \texttt{`TOGGLE\_TODO'}.

\begin{lstlisting}[language=TypeScript,style=acmcode, caption={Toy function with stringly comparisons},label={lst:lstex2},]
interface Action { type: string; payload?: any; }

function reducer(state: State, action: Action) {
  if (action.type === 'ADD_TODO') { return {...} };
  else if (action.type === 'REMOVE_TODO') { return { ... } }; 
  else if (action.type === 'TOGGLE_TODO') { return { ... }; }
  return state;
}
\end{lstlisting}

The goal of Discriminated Union is to formalize string-based state/type discrimination, preventing semantic drift and technical debt.

\textbf{(3) Union Alias:}
The Scope refers to locations in the code with semantically-related families of strings.
The Spec states that a union type containing the set of strings should exist.
Further, sites that produce/consume the strings must be annotated with the type and not rely on hard-coded strings.
For example, in `stringly-typed' APIs, promote a union (e.g., \texttt{type MessageType = 'info' | 'warning' | 'error'}) and constrain call sites (e.g., \texttt{processMessage(type: MessageType, data: any)}).
The goal of Union Alias is to centralize \\semantically-related literals used in similar contexts into a union type to prevent future semantic drift and duplication.

\textbf{(4) \texttt{satisfies} Guard:} 
The Scope refers to an object literal $M$ intended as a total mapping from a finite set of keys $(K)$ to values $(V)$ (e.g., HTTP methods $\rightarrow$ handlers).
The Spec states that the object $M$ has \textbf{\texttt{satisfies Record<K, V>}} and compiles successfully.
The goal of \textbf{\texttt{satisfies}} Guard is to ensure that finite-key objects have complete key coverage and correct values at runtime.

\emph{\textbf{Experiments}} We created a small proof-of-concept implementation written in TypeScript of the Type III side-car architecture (Figure~\ref{fig:sidecar}), with our previously outlined Type III component.
The autoformalization component uses LLM-based analysis to instantiate templates (1)-(4) as concrete specifications.
Currently, the Verification/Testing component of the side-car (Figure~\ref{fig:typethree}) uses lightweight syntactic checks and compilation success.
Future implementations will use more sophisticated code provenance for patterns that cannot be syntactically verified.
To implement the Type III system (Figure~\ref{fig:sidecar}), we integrated our proposed side-car with Claude Code, a popular vibe coding LLM agent, via hooks that run on code changes \cite{claude-code-hooks}. 
Results were promising; for example, the prototype system can emit the lines in \textbf{\textcolor{red}{red}} in Listing~\ref{lst:lstex}, flagging and fixing both bugs.
\section{Conclusion}

Vibe coding is inherently fragile. Prompt-by-prompt LLM edits inevitably collide with accumulating design constraints, leading to constraint-reconciliation decay, technical debt, and security failures.
In this position paper, we advocate for Vibe Reasoning (Type III systems): a developer-first integration of formal methods to address these fundamental limitations.
We invite the formal methods community to treat vibe coding's current pain points as fertile ground for developing the next generation of developer-centric verification techniques, and we hope our paper shines a light on the challenges and opportunities.

\onecolumn \begin{multicols}{2}

\bibliographystyle{ACM-Reference-Format}
\bibliography{sample-base}

\end{multicols}
\end{document}